\documentclass[usenatbib]{mn2e}
\usepackage{epsfig}
\usepackage{rotating}
\usepackage{multirow}
\usepackage{longtable}
\usepackage{breqn}
\usepackage{amssymb}

\bibpunct[, ]{(}{)}{;}{a}{}{,}

\title[J1649+26: A Spiral DRAGN]{J1649+2635: A Grand-Design Spiral with a Large Double-Lobed Radio Source}

\author[Mao et al.]{Minnie~Y. Mao$^{1}$\thanks{e-mail: mmao@nrao.edu}, Frazer Owen$^{1}$, Ryan Duffin$^{1,2}$, Bill Keel$^{3}$, Mark Lacy$^{4}$,
\newauthor  Emmanuel Momjian$^{1}$, Glenn Morrison$^{5,6}$, Tony Mroczkowski$^{7,8}$, Susan Neff$^{9}$, 
\newauthor Ray~P. Norris$^{10}$, Henrique Schmitt$^{8}$, Vicki Toy$^{11}$, Sylvain Veilleux$^{11,12}$\\
$^{1}$National Radio Astronomy Observatory, PO Box O, Socorro, NM, 87801, USA\\
$^{2}$University of Virginia, Charlottesville, VA 22904, USA\\
$^{3}$Department of Physics and Astronomy, University of Alabama, Box 870324, Tuscaloosa, AL 35487, USA\\
$^{4}$National Radio Astronomy Observatory, 520 Edgemont Road, Charlottesville, VA 22903, USA\\
$^{5}$Institute for Astronomy, University of Hawaii, Manoa, Hawaii 96822-1897 USA\\
$^{6}$Canada-France-Hawaii Telescope Corp., Kamuela, Hawaii 96743-8432, USA\\
$^{7}$National Research Council Fellow, USA\\
$^{8}$Remote Sensing Division, Naval Research Laboratory, 4555 Overlook Avenue SW, Washington, DC 20375, USA\\
$^{9}$Laboratory for Observational Cosmology, NASA Goddard Space Flight Center, Greenbelt, MD 20771, USA\\
$^{10}$CSIRO Astronomy and Space Science, PO Box 76, Epping, NSW, 1710, Australia\\
$^{11}$Department of Astronomy, University of Maryland, College Park, MD 20742-2421, USA\\
$^{12}$Joint Space-Science Institute, University of Maryland, College Park, MD 20742, USA\\
}

\begin{document}

%\date{2014}

\pagerange{\pageref{firstpage}--\pageref{lastpage}} \pubyear{2014}

\maketitle

\label{firstpage}

\begin{abstract}
We report the discovery of a grand-design spiral galaxy associated with a double-lobed radio source. J1649+2635 (z = 0.0545) is a red spiral galaxy with a prominent bulge that it is associated with a L$_{1.4\,{\rm GHz}}\sim$10$^{24}$\,W\,Hz$^{-1}$ double-lobed radio source that spans almost 100\,kpc. J1649+2635 has a black hole mass of M$_{\rm BH} \sim$ 3--7 $\times$ 10$^8$\,M$_{\odot}$ and SFR\,$\sim$ 0.26 -- 2.6\,M$_{\odot}$\,year$^{-1}$. The galaxy hosts a $\sim$96\,kpc diffuse optical halo, which is unprecedented for spiral galaxies. We find that J1649+2635 resides in an overdense environment with a mass of M$_{dyn} = 7.7^{+7.9}_{-4.3} \times 10^{13}$\,M$_{\odot}$, likely a galaxy group below the detection threshold of the ROSAT All-Sky Survey. We suggest one possible scenario for the association of double-lobed radio emission from J1649+2635 is that the source may be similar to a Seyfert galaxy, located in a  denser-than-normal environment. The study of spiral galaxies that host large-scale radio emission is important because although rare in the local Universe, these sources may be more common at high-redshifts. 

\end{abstract}


\section{Introduction}
\label{intro}
In the local Universe, powerful double-lobed radio sources are almost always associated with elliptical galaxies \citep[e.g.][]{Matthews64, Wilson95, Urry95, Best05}. This association is consistent with leading galaxy formation models \citep[e.g.][]{Hopkins06,Hopkins08}. 

It has long been postulated that elliptical galaxies form through the mergers of smaller galaxies \citep{Toomre72, Toomre77}. The merger of two gas-rich galaxies generates both star-formation and AGN (active galactic nucleus) activity. The merger triggers a powerful nuclear starburst \citep[e.g.][]{Armus87, Sanders88}, which generates strong starburst-driven winds, \citep{Rupke05, Veilleux05, Heckman00}. These winds will eventually blow away the enshrouding dust and lay bare the quasar core, depleting the dust and gas to form an elliptical galaxy \citep{Genzel01, Veilleux02, Tacconi02, Dasyra06a, Dasyra06b}. Although there are both starburst and quasar driven winds, quasar driven winds tend to occur at a later merger stage \citep[e.g.][]{Veilleux13, Rupke13}. At the same time, the merger fuels a pre-existing quiescent black hole \citep[i.e. radiatively efficient ``quasar-mode'' accretion,][]{Croton06}. A good example of this process in action is the ULIRG F00183-7111 \citep{Norris12, Mao14} where a radio-loud AGN has turned on, but the star-formation is still ongoing. This activity ceases when the fuel supply to the central regions is exhausted, and most of the remaining gas is expelled, starving both the active galactic nucleus (AGN) and the star-forming activity. Thereafter, the black hole is being fed more slowly (i.e. radiatively inefficient ``radio-mode'' accretion). That is to say, the merger of two gas-rich spiral forms an elliptical galaxy whilst launching radio-loud jets from the central AGN \citep[e.g][]{Chiaberge11}. The kpc-scale double-lobed sources associated with AGN are typically L$_{1.4\,GHz}$$\gtrsim$10$^{24}$\,W\,Hz$^{-1}$.

It is not uncommon for spiral galaxies with AGN to host low-power (L$_{1.4\,GHz}\sim$10$^{22}$\,W\,Hz$^{-1}$) radio jets, as in Seyfert galaxies. However the majority of these radio sources are sub-kpc-scale, and confined within the optical galaxy \citep[e.g][]{Ulvestad89, Kinney00, Schmitt01, Momjian03}. Although accretion of matter on to a central black hole can produce an AGN, a merger may be required to launch kpc-scale jets \citep{Chiaberge11}.

The process theorised to account for kpc-scale double-lobed radio sources requires the merger of two gas-rich galaxies, which eventually forms an elliptical. This model does not result in a gas-rich spiral hosting a powerful radio-emitting AGN.

\subsection{Radio Galaxies that are hosted by Spirals}

Powerful radio sources that are hosted by spiral and disk-dominated galaxies are known to exist at low-redshift, although to-date there are only a few tens of examples \citep[e.g][]{Morganti11}, most of which are hosted by disk-dominated or lenticular galaxies \citep[e.g. NGC612,][]{Ekers78}. The few examples of these have been studied in detail. Table \ref{spirals} gives a complete-to-date list of \emph{spiral} galaxies that host large double-lobed radio sources\footnote{Radio Galaxy Zoo (Banfield et al. in prep) enables citizen scientists to find more of these sources, and has so far identified tens of candidates.}. 

\begin{table*}
\begin{center}
\caption{Table of spiral galaxies associated with large double-lobed radio sources. We include Speca in this Table as it is denoted as a spiral host in \citet{Hota11}, however we note that while disk structure is clearly apparent in this galaxy, the spiral structure is somewhat contentious. }\label{spirals}
\begin{tabular}{llllll}
\\
\hline
ID & z & FR & L$_{1.4\,GHz}$ & Environment & Reference\\
\hline
0313-192 & 0.067 & FRI & $1.0 \times 10^{24}$\,W\,Hz$^{-1}$ & Abell 428 & \citet{Ledlow98}\\
Speca & 0.138 & FRII & $7.0 \times 10^{24}$\,W\,Hz$^{-1}$ & BCG in MaxBCG J212.45357-03.04237 & \citet{Hota11}\\ 
J2345-0449 & 0.076 & FRII & $2.5 \times 10^{24}$\,W\,Hz$^{-1}$ & Field or peripheral member of RBS2042 & \citet{Bagchi14}\\
J1649+26 & 0.055 & FRII & $1.0 \times 10^{24}$\,W\,Hz$^{-1}$ & BCG in Zwicky Cluster 1648.3+2632 & this work\\
\hline
\end{tabular}
\end{center}
\end{table*}

0313-192 was the first confirmed double-lobed radio source hosted by a spiral galaxy \citep{Ledlow98, Ledlow01, Keel06}\footnote{The first mention in the literature of a spiral hosted double-lobed radio source was PKS 0400-181. Higher quality CCD images showed it to be a chance alignment \citep{Shaver83}.}. 0313-192 is an edge-on spiral galaxy (Hubble type Sb) that hosts a 360 kpc double-lobed radio source. The galaxy, at z\,=\,0.067, resides in the poor cluster Abell 428, which is a group of groups. The radio source has a 1.4\,GHz integrated flux density of 98\,mJy, which translates to a total power of L$_{1.4\,{\rm GHz}}\sim$10$^{24}$\,W\,Hz$^{-1}$. This is consistent with the morphological class FRI \citep{Fanaroff74}. \citet{Keel06} suggest the unusual nature of 0313-192 may be attributed to the combination of a number of factors. Firstly, the over-luminous bulge implies that the central black hole is unusually massive, so jets emanating from it may travel faster than those seen in Seyfert galaxies. Secondly, the projected orientation of the radio lobes is almost exactly perpendicular to the disk of the host spiral galaxy in the plane of the sky. The combination of fast jets and less interstellar medium (ISM) for them to burrow through may allow the presence of radio lobes. Finally, the presence of a warped stellar disk suggests a minor merger has occurred in the galaxy's past.

Speca (J1409-0302) was the second spiral galaxy found to host a large double-lobed radio source \citep{Hota11}. The host galaxy, at z\,=\,0.138, is optically red yet also shows signs of recent star formation. It is the brightest cluster galaxy (BCG) in the galaxy cluster MaxBCG J212.45357-03.04237. The radio emission spans 307\,kpc and appears to be episodic. Furthermore, there is evidence for old ``relic'' radio lobes that span $\sim$1.3\,Mpc, suggesting this radio source may be a giant radio galaxy\footnote{Giant radio sources are defined to be those with a projected linear size $>$1 Mpc \citep{Ishwara99}.}. \citet{Hota11} suggest that due to the unrelaxed nature of the galaxy cluster, and the evidence for recent star-formation, Speca may be accreting cold gas from the ICM. This would remove the need for a merger to feed the black hole.

J2345-0449 was very recently discovered as a giant radio source hosted by a spiral galaxy \citep{Bagchi14}. The host galaxy, at z\,=\,0.0755, appears to reside in an isolated environment, however a massive galaxy cluster RBS 2042, at the same redshift, is located $\sim$2.8\,Mpc away. The radio emission spans $\sim$1.63\,Mpc, has an FRII morphology and appears episodic, similar to Speca. Also similar to Speca, the authors suggest that a merger is not required to fuel the AGN, instead proposing a scenario involving a fast-spinning black hole and relatively inefficient accretion. 

\subsection{J1649+2635}

To further our understanding of spiral galaxies that host large-scale radio lobes, we started a project to systematically search for spiral galaxies associated with large-scale radio lobes \citep{Duffin14}. 

We cross-matched morphologically classified optical galaxies from Galaxy Zoo \citep{Lintott08} with the Unified Radio Catalog \citep{Kimball08}, which contains data from both the Faint Images of the Radio Sky at Twenty-Centimeters \citep[FIRST, ][]{Becker95} and the NRAO VLA Sky-Survey \citep[NVSS, ][]{Condon98}. The full Galaxy Zoo catalogue contains morphological data for 667,944 galaxies.  We used the ``superclean'' subset of this dataset, which contains 65,492 galaxies. The ``superclean'' sample requires consistent identification of the galaxies as spiral or elliptical by at least 95\% of Galaxy Zoo users. Extended radio sources were identified by comparing the lower resolution NVSS flux density measurements ($\theta \sim 45$'') to those of FIRST ($\theta \sim 5$''). We applied a radio luminosity cutoff L$_{1.4\,GHz}$ = $10^{23}$\,W\,Hz$^{-1}$ to filter out star-forming galaxies.

The resulting catalogue contained three sources that were identified as spiral galaxies in the ``superclean'' sample from Galaxy Zoo, and had extended radio emission with L$_{1.4\,GHz} \ge 10^{23}$\,W\,Hz$^{-1}$. One was a chance-alignment and another was a nearby star-forming galaxy. The third galaxy, SDSS J164924.01+263502.5 (henceforth J1649+2635), hosts a powerful, double-lobed radio source and is the subject of this paper\footnote{96.8\% of Galaxy Zoo users determined this galaxy to be a spiral and 90\% of those users agreed on the rotational direction of the arms.}.

In this paper we report the discovery of J1649+2635, a double-lobed radio source whose host galaxy has a spiral morphology. The spiral nature was independently ascertained, prior to any association with a powerful double-lobed radio source. This paper uses H$_0$ = 71 km s$^{-1}$ Mpc$^{-1}$, $\Omega$$_M$ = 0.27 and $\Omega$$_\Lambda$ = 0.73 \citep{Spergel03} and the web-based calculator of \citet{Wright06} to estimate the physical parameters. At the redshift of J1649+2635 (z\,=\,0.0545), one arcsecond corresponds to 1.046\,kpc. 

 \section{Data}\label{data}

\subsection{Archival Data}
\begin{figure*}
\begin{center}
\begin{tabular}{cc}
\includegraphics[angle=-90, scale=0.32]{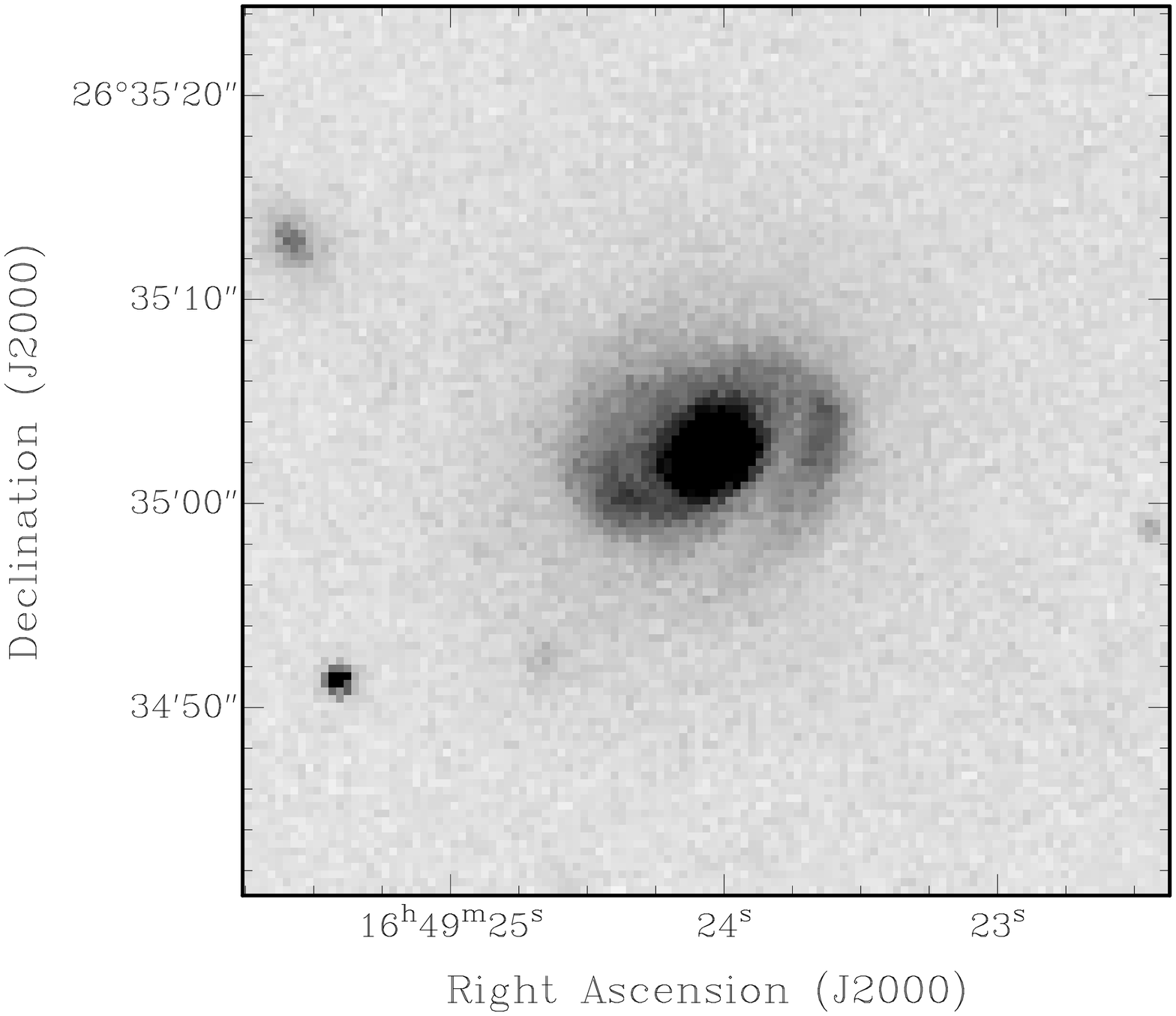}&\includegraphics[angle=-90, scale=0.32]{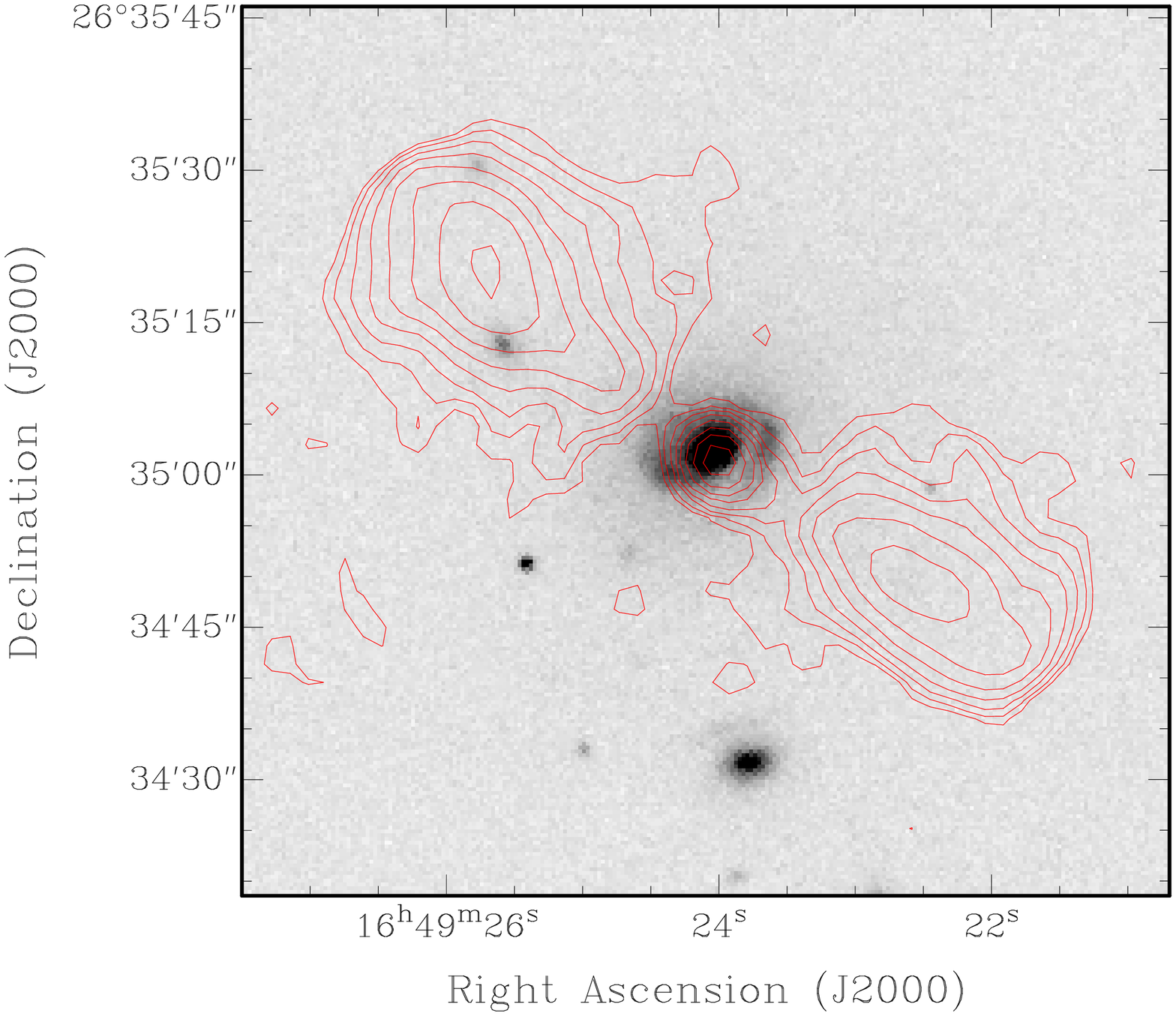}\\
\end{tabular}
\end{center}
\caption{\textbf{Left:} SDSS DR9 g-band. \textbf{Right:} 1.4\,GHz FIRST radio contours overlaid on SDSS DR9 g-band image. Contours start at 3$\times$ 0.15\,mJy (rms of FIRST) and increase by factors of $\sqrt2$.   }\label{image}
\end{figure*}

J1649+2635 was classified as a spiral galaxy in the Galaxy Zoo project, which utilises the Sloan Digital Sky Survey (SDSS). The spiral structure is evident in the left panel of Figure \ref{image}. The galaxy has a g-band magnitude of 15.50 \citep{Abazajian09}. 

The radio source associated with J1649+2635 has been identified in both the NVSS and FIRST surveys at 1.4\,GHz. The NVSS image of the source is slightly resolved and has a total integrated flux density of 157\,mJy, which translates to a radio power of L$_{1.4 {\rm GHz}}$ = 1.1\,$\times$\,10$^{24}$\,W\,Hz$^{-1}$.  The FIRST image of the source clearly shows a double-lobed structure and has a total integrated flux density of 108\,mJy at 1.4\,GHz (Figure \ref{image}).  This discrepancy may suggest that some larger-scale emission is detected in NVSS but not in FIRST. The total integrated flux density of J1649+2635 at various frequencies is presented in Table \ref{1649data}. Due to difficulties determining absolute radio flux density errors, we estimate a conservative error of 10\% in all calculations. 

%diffuse emission is resolved out in the FIRST data. 

\begin{table*}
\begin{center}
\caption{Properties of J1649+2635.  WISE magnitudes are in the Vega system. All radio flux densities, with the exception of FIRST, give the total integrated flux. }\label{1649data}
\begin{tabular}{lllll}
\\
\hline
RA (J2000)& 16:49:24.01 & &&\\
Dec (J2000)& 26:35:02.6 & &&\\
Redshift & 0.0545 & &&\citet{Ahn12}\\
\hline
Survey & Band & Flux Density & Magnitude & Reference\\
\hline
GALEX NUV & 2267\AA & 26.36\,$\mu$Jy & & \citet{Budavari09}\\
SDSS & u & & 17.65 & \citet{Abazajian09}\\
SDSS & g & & 15.50 & \citet{Abazajian09}\\
SDSS & r & & 14.48 & \citet{Abazajian09}\\
SDSS & i & & 13.97 & \citet{Abazajian09}\\
SDSS & z & & 13.58 & \citet{Abazajian09}\\
W1 & 3.4\,$\mu$m & &11.89 & \citet{Wright10}\\
W2 & 4.6\,$\mu$m & &11.87 & \citet{Wright10} \\
W3 & 12\,$\mu$m & &9.74 & \citet{Wright10}\\
W4 & 22\,$\mu$m & &8.04 & \citet{Wright10}\\
GB6 & 4.85\,GHz & 63\,mJy & &\citet{Gregory91}\\
NVSS & 1.4\,GHz & 157\,mJy& & \citet{Condon98}\\
FIRST & 1.4\,GHz & 102\,mJy& & \citet{Becker95}\\
B2 & 408\,MHz & 331\,mJy & &\citet{Colla72}\\
VLSSr & 74\,MHz & 640\,mJy && \citet{Lane12}\\
\hline

\end{tabular}
\end{center}
\end{table*}

\subsection{Optical Data from the Discovery Channel Telescope}

J1649+2635 was observed using the 4.3 m Discovery Channel Telescope \citep[DCT,][]{Bida12, Levine12, DeGroff14} equipped with the $12.5' \times 12.5'$ field-of-view Large Monolithic Imager (LMI) located in Arizona, USA.  The galaxy was observed in SDSS r-band on 2014 June 8 for $4 \times 5$ minute exposures with seeing of $\sim$1.7 arcseconds and on 2014 June 9 for $5 \times 5$ minute exposures with seeing of $\sim$1.2 arcseconds for a total of $9 \times 5$ minute dithered exposures. 

The individual observations were reduced using a custom IRAF software pipeline.  The individual frames were astrometrically corrected using the Scamp software package and coadded using Swarp \citep{Bertin02}.

\section{Results and Discussion}

\subsection{The Optical Host}
J1649+2635 is identified as a spiral galaxy at a redshift z\,=\,0.0545 \citep{Ahn12}. The spiral arms are clearly apparent in Figure \ref{image} (Left), as is a very luminous bulge. Morphologically, there is some evidence for a bar structure, and perhaps a hint of a dust lane to the South-West of the bulge. This may suggest the galaxy is tilted with its North side towards us. The galaxy fits the description of a ``grand-design spiral''. 

The spectrum of J1649+2635, available from the SDSS website, is similar to that of a standard early-type galaxy. However, the SDSS fibres are 3\,arcsec in diameter (corresponding to $\sim$3.2\,kpc at z\,=\,0.0545) so the SDSS spectrum is only of the bulge of the galaxy, whose spectrum is that of an early-type galaxy \citep[see e.g.][]{Mauch07, Mao12}. 

The galaxy is unusually red in colour for a spiral galaxy, which is inconsistent with ``normal'' spirals that are typically blue in colour \citep[e.g.][]{DeVaucouleurs61}. The Galaxy Zoo project discovered the ``red spiral'' class of objects \citep{Bamford09} and \citet{Masters10} find that just 6 per cent of face-on spiral galaxies are red in colour. Red spirals are found to have similar star-formation histories to blue spirals until 500\,Myr ago, after which red spirals have up to 3$\times$ less star formation \citep{Tojeiro13}. \citet{Schawinski14} suggest that the star-formation of red spirals has been quenched by the cool gas flow being cut-off by some mechanism such as a merger. Red spirals are preferentially found in intermediate density regimes \citep{Bamford09, Masters10}. Using the criterion (g-r)\,$> 0.63-0.02(\rm{M_r} + 20)$ \citep{Masters10}, J1649+2635 is clearly a red spiral (M$_{\rm r} = -22.73$, g-r\,=\,1.02). However, the galaxy is not classified as a red spiral in \citet{Masters10}, as they impose additional criteria to select against spirals dominated by light from the bulge. 

\begin{figure}
\begin{center}
\includegraphics[angle=-90, scale=0.32]{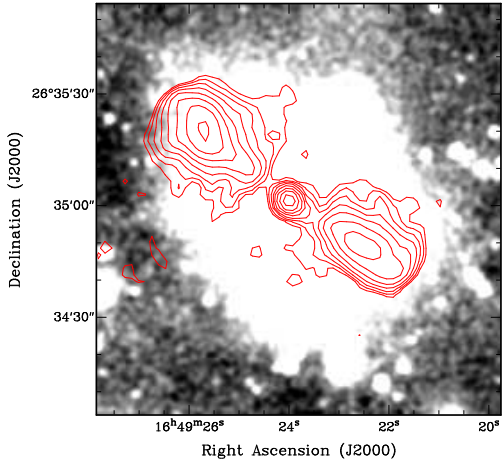}
\end{center}
\caption{DCT r-band image of J1649+2635 smoothed with the saturation turned up to display the diffuse halo. The FIRST radio contours have been overlaid for comparison. The contour levels are as for Figure \ref{image}. }\label{dct_smooth}

\end{figure}

\begin{figure*}[tbh]
\begin{center}
\includegraphics[angle=0, scale=0.8]{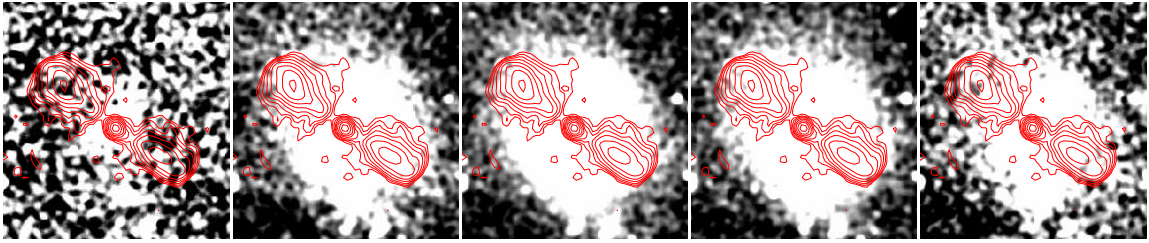}
\end{center}
\caption{SDSS Images of J1649+2635 smoothed and with the saturation turned up to display the diffuse optical halo. The panels from left to right are the u,g,r,i and z images respectively. The FIRST radio contours have been overlaid for comparison. The contour levels are as for Figure \ref{image}.  }\label{halo}
\end{figure*}

There is evidence for a diffuse optical halo spanning $\sim$90'', which corresponds to 96\,kpc at z = 0.0545 (Figure \ref{dct_smooth}). The halo is present in all SDSS filters, except for the u-band filter (Figure \ref{halo}). This is consistent with an older stellar population occupying the halo, but its absence in the u-band may be due to the decreased sensitivity of the shortest-wavelength band. 

\begin{figure}
\begin{center}
\includegraphics[angle=90, scale=0.5]{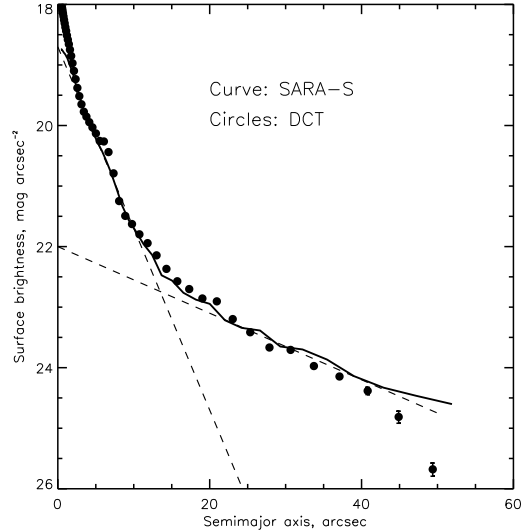}
\end{center}
\caption{Surface-brightness profile of J1649+2635 from the DCT r-band data. The profile may be fit by two exponentials, and the extended component is visible to a radius of $\sim50''$, and may extend even further. The downturn at the $\sim45''$ is likely due to the low signal-to-noise. }\label{billfig}
\end{figure}

The surface-brightness profile for J1649+2635 may be fit by two exponentials, corresponding to the spiral disk and the extended halo (Figure \ref{billfig}). The integrated model profile of this system is r=13.44, which suggests the integrated halo brightness if comparable to that of the spiral disk. The optical halo  extends to a radius of $\sim45''$. From the profile, it is possible that the optical halo may extend even further, however deeper images are required to confirm this.

Massive optical halos of this nature are only found surrounding the centrally dominant cD galaxies of clusters. There are no known examples of spiral galaxies hosting giant optical halos of this extent. That is to say, J1649+2635 is \emph{the first instance of a grand-design spiral hosting a large diffuse optical halo}.

The presence of a halo strongly suggests that J1649+2635 sits near the bottom of a gravitational potential well. As the halos surrounding cD galaxies may result from multiple mergers believed to form these galaxies, this halo may also suggest a period of disruption or merger for J1649+2635, although this invokes the question of whether spiral structure may be maintained post-merger. 

The extent of the radio source (discussed below) is comparable to that of the halo. The similarity of the size and shape of the optical halo to the radio source may indicate that gas trapped in the galaxy's potential well is the medium which is confining the radio source and responsible for the radio emission being visible.

\subsection{The Radio Source}

The radio source associated with J1649+2635 has a total radio luminosity of L$_{1.4 GHz}$ = 1.03\,$\times$\,10$^{24}$\,W\,Hz$^{-1}$ (using the NVSS). The total integrated flux of this source is $\sim$30 per cent larger in the lower-resolution NVSS survey than the FIRST survey. This may be due to the presence of diffuse larger-scale radio emission that is detected in the NVSS but not in FIRST due to a lack of shorter baselines. Interestingly, both Speca and J2345-0449 show evidence for diffuse megaparsec-scale radio structures and appear episodic. We do not detect any such structure in J1649+2635 in archival low-frequency observations (VLSSr, B2), but this is likely due to their shallow-nature. We are in the process of obtaining deeper lower-frequency radio data to determine whether J1649+2635 also displays diffuse, episodic characteristics.

%diffuse emission may have been resolved out in the higher resolution observations. 
The radio lobes subtend approximately 80$''$, giving an end-to-end length of $\sim$~85\,kpc. Figure \ref{image} (Right) clearly demonstrates the existence of an FRII double-lobed structure. The core has a peak flux of $\sim 6$\,mJy\,beam$^{-1}$ at 1.4\,GHz. %The core and lobes have similar surface brightnesses, suggesting an FRI/FRII transitional morphology.

\begin{figure}
\begin{center}
\includegraphics[angle=0, scale=0.5]{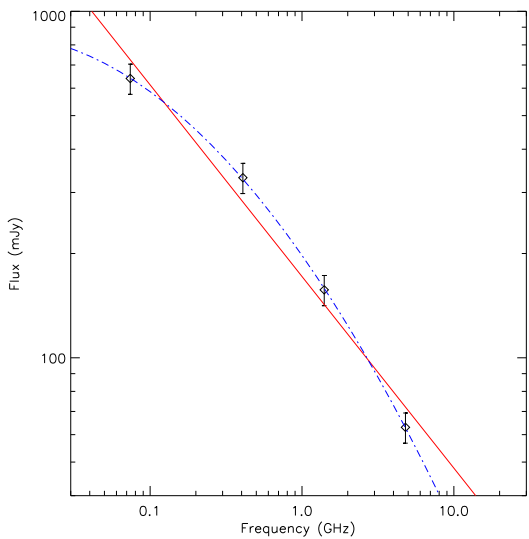}
\end{center}
\caption{Spectral index plot for the integrated radio emission from J1649+2635, with 10\% errors. The solid red line is the linear least-squares fit (log(S)$= -0.55$log($\nu)+2.23$) and the blue dot-dashed line is a polynomial least-squares fit (log(S)$= 2.30 + -0.63$log($\nu) -0.15$(log($\nu$))$^2$). }\label{si}
\end{figure}

Figure \ref{si} plots the spectral index of J1649+2635 using integrated flux densities over two decades of frequency. A linear least-squares fit finds the spectral index to be $\alpha \sim -0.55$, where  $S \propto \nu^{\alpha}$. This is flatter than normal synchron emission ( $\alpha \sim -0.7$). When the 74\,MHz VLSSr data are omitted from the spectral index calculation, $\alpha \sim -0.67$. We are in the process of obtaining higher resolution radio data, which will be presented in a future publication\footnote{Preliminary results from these data suggest the radio source to have $\alpha^{9GHz}_{1.4GHz} \sim -0.02$ at its core and $\alpha^{9GHz}_{1.4GHz} \sim = -0.58$ in its lobes.}. We note here that the spectrum is better fit by a second-order polynomial. 

\subsection{Evidence for star-formation }
J1649+2635 was detected in the Medium Imaging Survey by GALEX \citep{Martin05} at NUV wavelengths and not observed at FUV wavelengths\footnote{J1649+2635 was observed at both NUV and FUV wavelengths as part of the All-sky Imaging Survey (AIS), but was only detected at NUV.}. The source is slightly resolved and has a flux density of 26.37\,$\mu$Jy, corresponding to L$_{NUV}$ = 1.84\,$\times$\,10$^{20}$\,W\,Hz$^{-1}$. From \citet{Kennicutt98}, we estimate a lower limit to the star-formation rate (SFR) of 0.26\,M$_{\odot}$\,year$^{-1}$. However, as noted by \citet{Kennicutt98}, the true SFR may be over an order of magnitude higher because of uncertainty in the correction for extinction by dust. 

The source is detected in all four WISE \citep{Wright10} bands. Using colour-colour relationships \citep[e.g][]{Lacy04,Yan13}, the measured [W1]-[W2] = 0.02 is clearly classified as star-formation (with [W1-W2]$>$0.8 being AGN). We use the 22\,$\mu$m data\footnote{This is just an approximation as we assume this value is sufficiently close to the 24\,$\mu$m Spitzer band.} to estimate the SFR using the relations from \citet{Galametz13} and \citet{Kennicutt98}, and calculate SFR $\sim$ 1.89\,M$_{\odot}$\,year$^{-1}$. Taking into account the large uncertainties associated with these SFR estimates, we estimate SFR\,$\sim$\,0.26 -- 2.6\,M$_{\odot}$\,year$^{-1}$.

\subsection{Black Hole Mass Estimate for J1649+2635}
We used available archival data to estimate the mass of the black hole in J1649+2635 using the black hole mass -- bulge luminosity relationship \citep{Kormendy95, Magorrian98, Kormendy13}. 

For J1649+2635, the absolute bulge magnitudes are M$_{\rm g}$\,=\,-21.19, M$_{\rm r}$\,=\,-22.17 and M$_{\rm V}$ = -21.77. The values are from from \citet{Simard11}, who performed bulge-disk decompositions on galaxies in SDSS Data Release 7. Using the values of  $\alpha$ = 9.23 and $\beta$ = 1.11 \citep[for early-type galaxies,][]{McConnell13}, we estimate a black hole mass for J1649+2635 of M$_{\rm BH}$ = 6.7 $\times$ 10$^8$\,M$_{\odot}$. Using $\alpha$ = 9.10 and $\beta$ = 0.98 \citep[for L $\le 10^{10.8}$,][]{McConnell13}, we estimate M$_{\rm BH}$ = 5.6 $\times$ 10$^8$\,M$_{\odot}$. For a sanity check we also used the parameter estimates from \citet{Bentz09} and find M$_{\rm BH}$ = 3.1 $\times$ 10$^8$\,M$_{\odot}$. 

We therefore estimate the black hole mass to be M$_{\rm BH} \sim$ 3--7 $\times$ 10$^8$\,M$_{\odot}$. This is similar to that found for galaxies that host powerful radio AGN \citep[e.g][]{Chiaberge11}, and considerably larger than for spiral galaxies such as the Milky Way \citep[M$_{\rm BH} \sim 4 \times 10^6$\,M$_{\odot}$,][]{Ghez08}. These values are similar to the black hole mass estimates for 0313-192 \citep{Keel06} and J2345-0449 \citep{Bagchi14}. 

It has been empirically determined that radio-loud AGNs invariably host black holes with masses of  $\ge 10^8$\,M$_{\odot}$ \citep[e.g.][]{Chiaberge05}. \citep{Chiaberge11} suggest that radio-loud AGN require both a black hole mass $\ge 10^8$\,M$_{\odot}$, and significant spin of the black hole. This may be achieved through mergers.

\subsection{The Environment of J1649+2635}
\citet{Yoon08} previously identified a cluster near the position of J1649+2635, with the ID Zw8068, likely from the Zwicky catalogue \citep{Zwicky63}, where the cluster is identified as 1648.3+2632.

\begin{figure*}
\begin{center}
\begin{tabular}{cc}
\includegraphics[angle=0, scale=0.5]{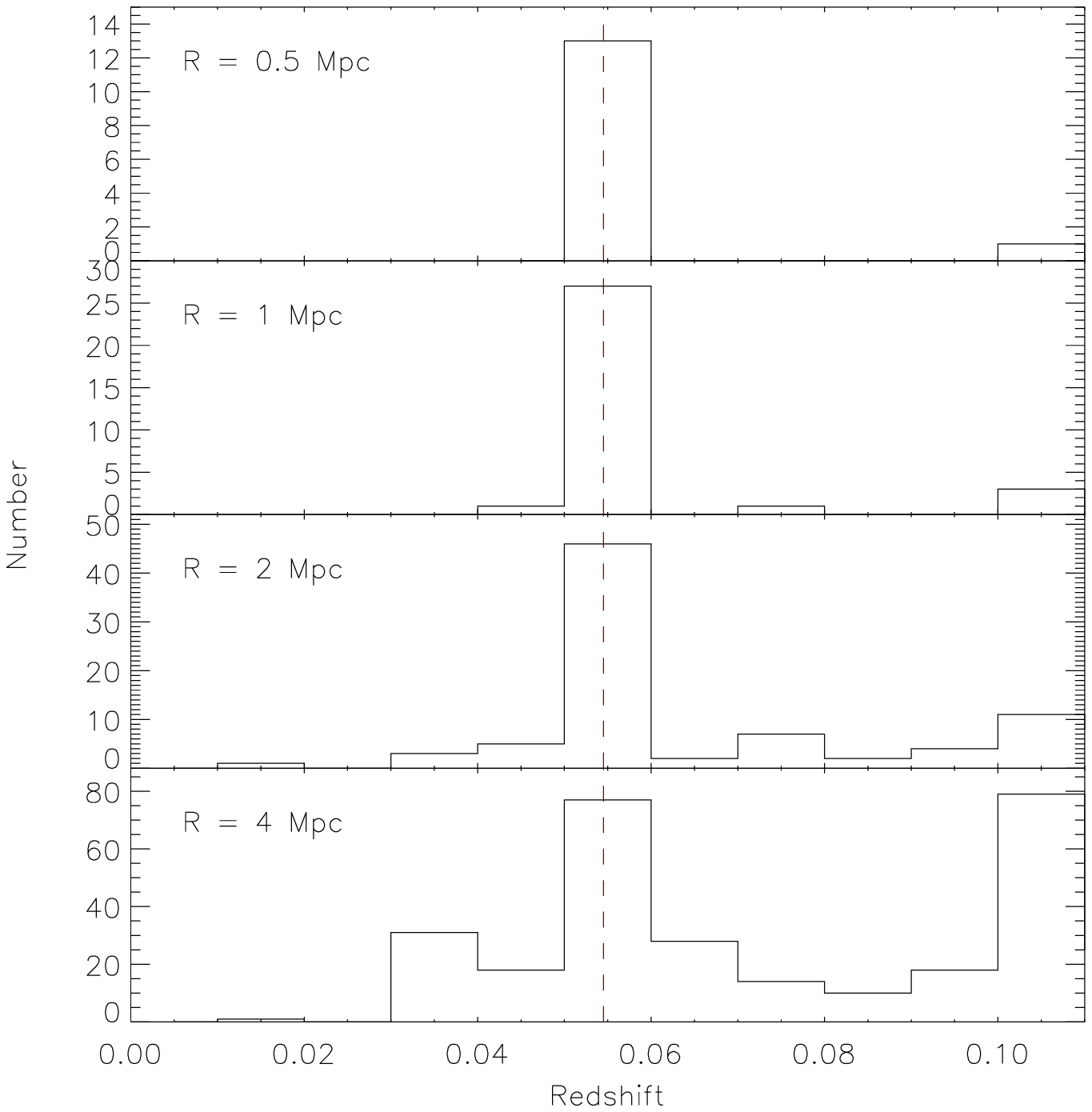} & \includegraphics[angle=0, scale=0.5]{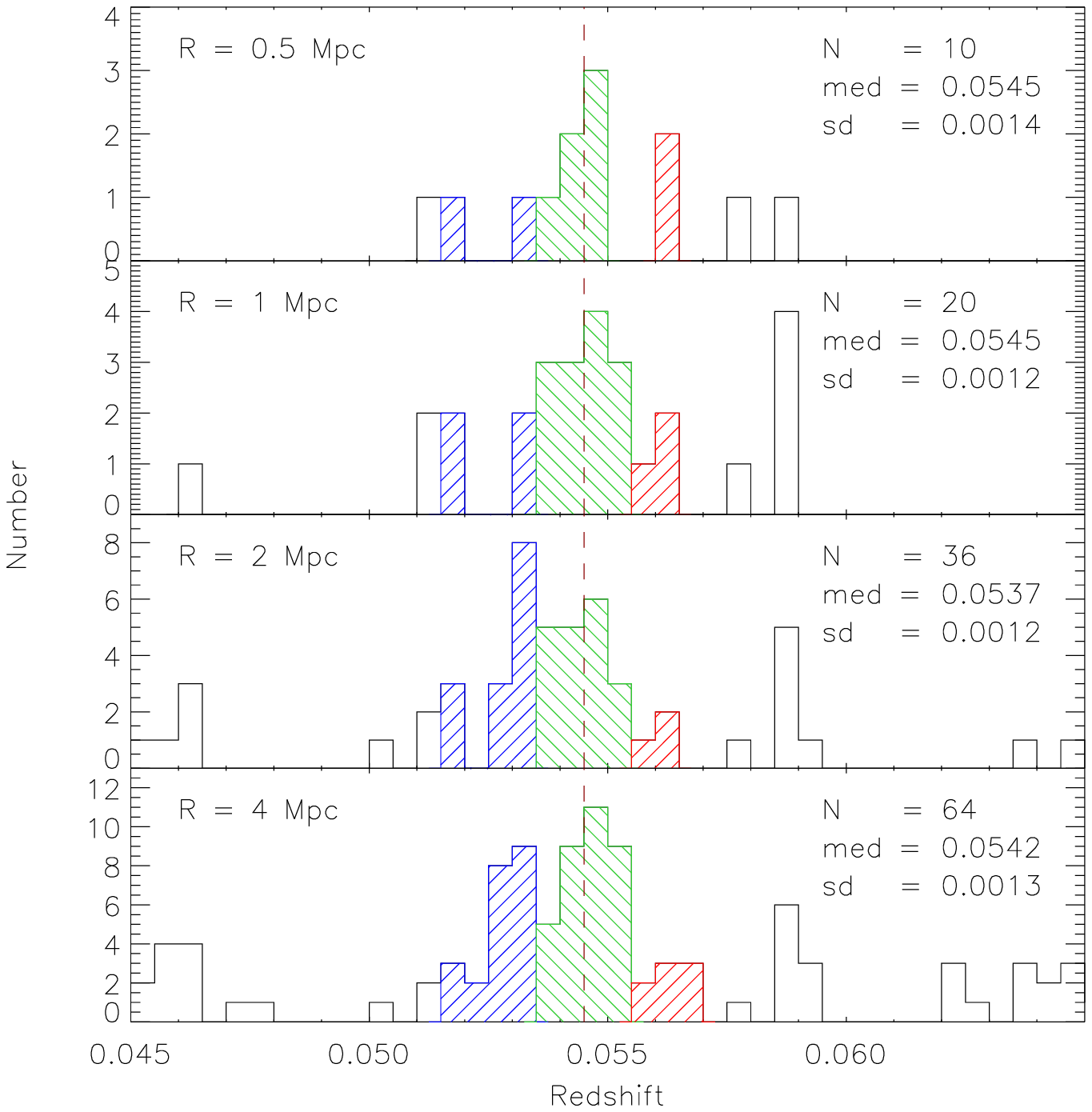}\\
\end{tabular}
\end{center}
\caption{\textbf{Left}: The redshift histogram for all sources at z\,$\le$\,0.11. The histogram binsize is $\Delta z = 0.01$, which corresponds to 3000\,km\,s$^{-1}$. The topmost panel displays the histogram for the region within a 0.5\,Mpc radius of J1649+2635 and the bottommost panel is for the region within a 4\,Mpc radius. The red dashed line marks the redshift of J1649+2635. \textbf{Right}: Redshift histograms for the same regions, but with a narrower redshift range. The histogram binsize is $\Delta z = 0.0005$, which corresponds to 150\,km\,s$^{-1}$. The blue, green, and red hatched region displays sources at $0.0515 \le$\,z\,$< 0.0535$, $0.0535 \le$\,z\,$< 0.0555$, and $0.0555 \le$\,z\,$< 0.0575$ respectively. The number of sources, median and standard deviation are computed from the coloured section of the histogram. }\label{hist}
\end{figure*}

Clusters of galaxies are typically a few Mpc in diameter, however as we have not yet identified the cluster centre, we investigate a larger area. We used the NASA Extragalactic Database (NED) to obtain spectroscopic redshifts in the immediate vicinity of J1649+2635. Within a 64.9 arcmin radius (4\,Mpc at z\,=\,0.0545), there were 590 galaxies with spectra, of which 276 were z\,$\le$\,0.11. The histogram of redshifts is plotted in Figure \ref{hist}. There is clearly a peak at the redshift of J1649+2635, indicating the presence of a group or cluster of galaxies. The right panel of Figure \ref{hist} displays the same histogram at a higher velocity resolution in the redshift range surrounding J1649+2635. Within a radius of R = 1\,Mpc, the velocity dispersion is 360\,km\,s$^{-1}$. This is consistent with poor clusters, or groups of galaxies in the local Universe \citep[e.g][]{Bahcall93}. There is some evidence that the redshift histogram is not well-represented by a Gaussian, especially at R $\ge$ 1\,Mpc, suggesting an unvirialised system. 

\begin{figure}
\begin{center}
\includegraphics[angle=-90, scale=0.35]{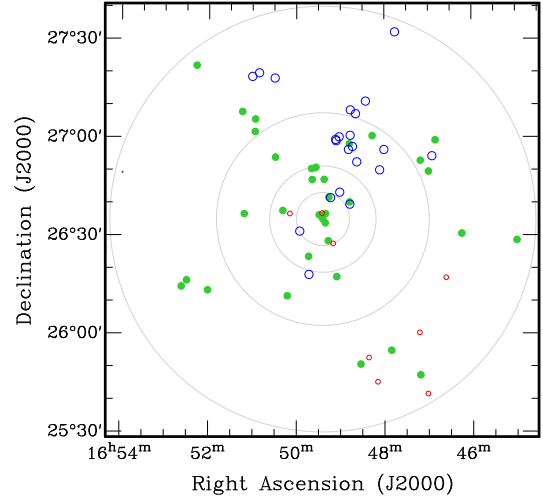}
\end{center}
\caption{The spatial distribution of galaxies around J1649+2635, that are within  $\Delta$z = 0.003 ($\Delta$cz = 900\,km\,s$^{-1}$) of z=0.0545, the redshift of J1649+2635. The colours correspond to the colour scheme of Figure \ref{hist}. The closest redshift range (blue) are larger open circles, the redshift range of J1649+2635 (green) are closed circles, and the most distant redshift range (red) are smaller open circles. The concentric grey circles represent 0.5, 1, 2, and 4\,Mpc radii respectively.}\label{radec}
\end{figure} 

Figure \ref{radec} plots the spatial location of the 64 galaxies that are within $\Delta$z = 0.003 ($\Delta$cz = 900\,km\,s$^{-1}$) of J1649+2635. There is a hint of a velocity gradient across the field, with the closest galaxies (blue larger open circles) predominately north of J1649+2635, and the more-distant galaxies (red smaller open circles) predominately south-west of J1649+2635. This may be further evidence for an unrelaxed system of galaxies. Perhaps the system is a group of groups, similar to the environment surrounding 0313-192. 

\begin{figure*}
\begin{center}
 \includegraphics[angle=0, scale=0.8]{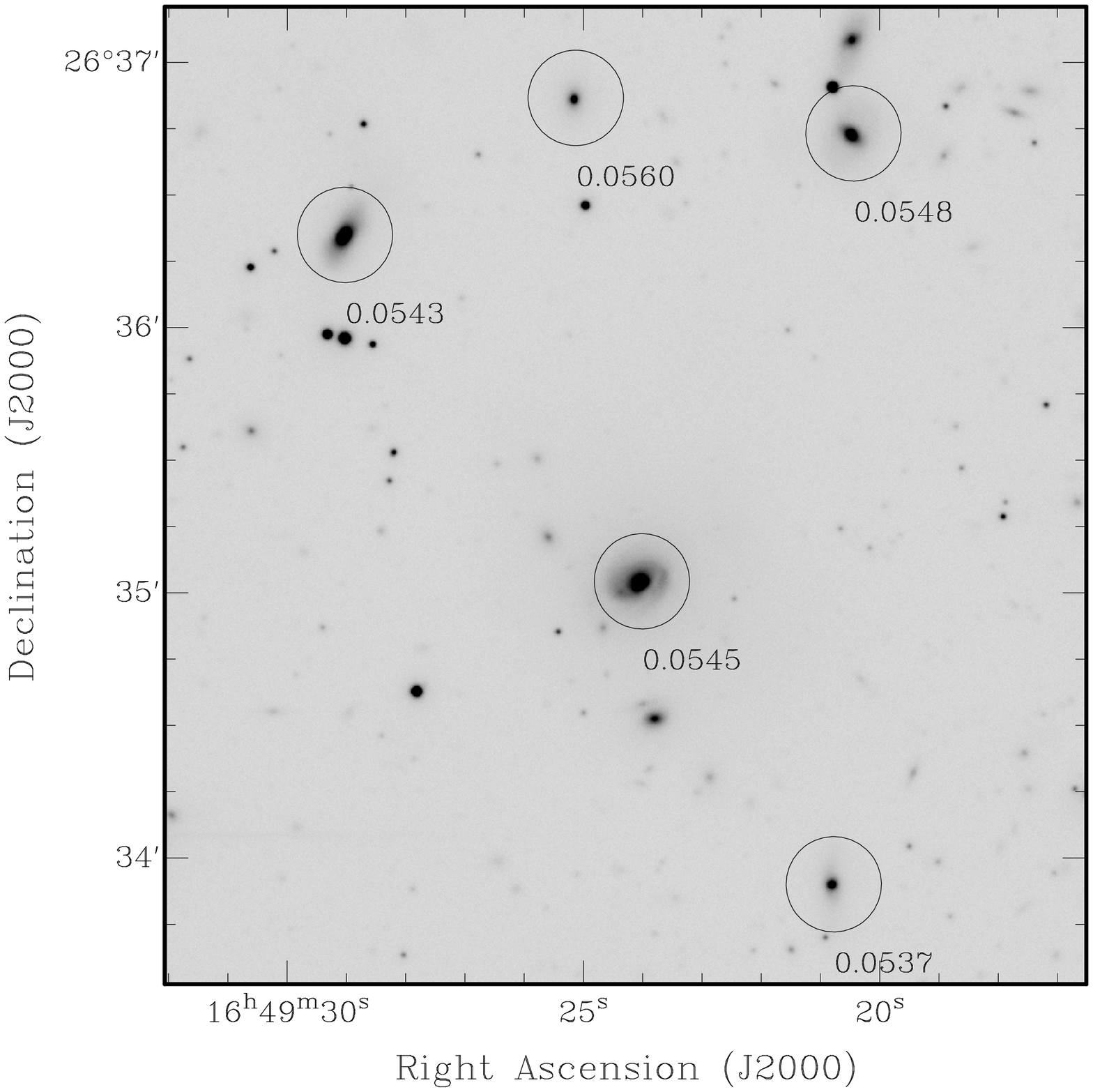}
\end{center}
\caption{DCT r-band image of the region surrounding J1649+2635. Galaxies that have spectroscopic redshifts within 900\,km\,s$^{-1}$ of J1649+2635's redshift are indicated by circles.  }\label{dct}
\end{figure*}

Groups contain higher fractions of disk galaxies than clusters \citep[e.g]{Dressler80}. Figure \ref{dct} shows the DCT image with the galaxies that are at the same redshift circled in black. All of these galaxies show evidence for disk-like structure. In contrast, the morphologies of cluster members are often ellipticals. This is further evidence that this system is lower density than a classical cluster of galaxies. However, the large optical halo around J1649+2635 is more reminiscent of a cluster environment. 

Using the 1D projected velocity dispersion of the system (Figure \ref{hist}), and following the relation from \citet{Saro13}, we compute the dynamical mass of the system to be M$_{dyn} = 7.7 \pm 1.7 \times 10^{13}$\,M$_{\odot}$, where the errors are due to galaxy membership within the aperture used. Further, \citet{Saro13} note that due to the difficulty in relating the projected velocity dispersion to the true 3D velocity dispersion, the systematic error may be up to a factor of $\sim$2 higher at low redshift. Consequently we estimate the dynamical mass of the system to be M$_{dyn} = 7.7^{+7.9}_{-4.3} \times 10^{13}$\,M$_{\odot}$. This is consistent with the virial masses of groups of galaxies \citep[e.g.][]{Connelly12}. We have examined archival \emph{ROSAT} All-Sky Survey (RASS) data, and have found no detection towards J1649+2635. This implies an upper limit to the X-ray luminosity of $2.12 \times 10^{43}$\,erg\,s$^{-1}$\,Hz$^{-1}$ \citep{Boehringer01}, and indicates an upper limit to the mass of $\sim 1 \times 10^{14}$\,M$_{\odot}$ \citep{Pratt09}, consistent with the value obtained using the velocity dispersion. 

We also note that J1649+2635 (m$_{\rm g}$ = 15.5) is one of the two brightest galaxies in this system. The other bright galaxy, SDSS J164933.52+265052.0 is at z = 0.0551 and is $\sim$1\,Mpc away. It is possible that this galaxy is the cD galaxy for a group or cluster that is in the process of interacting with a group or cluster that J1649+2635 resides in. There is no evidence for a larger-than-usual diffuse optical halo around the other bright galaxy. 

\begin{figure}
\begin{center}
\includegraphics[angle=0, scale=0.5]{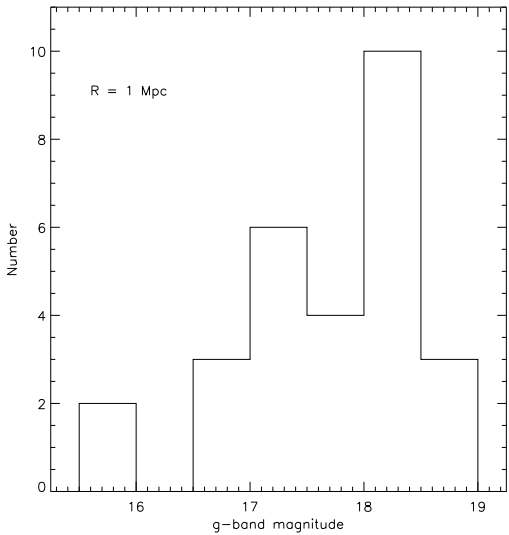}
\end{center}
\caption{Histogram of g-band magnitudes for all galaxies within a radius of 1\,Mpc of J1649+2635, and with 0.045 $\le$ z $<$ 0.065. The g-band magnitude of J1649+2635 is 15.5. }\label{maghist}
\end{figure} 

Given the disk-like morphologies of the galaxies in the vicinity of J1649+2635, and the dynamical mass estimate, we find the environment of J1649+2635 is more like a group of galaxies, than a cluster. This is also consistent with J1649+2635 being a red spiral, as red spirals are usually found in intermediate density environments. 

\subsection{Possible causes for the curious nature of J1649+2635}
J1649+2635 is clearly ``hybrid'' in nature; it is a star-forming spiral galaxy that has large-scale radio lobes. Although this is generally unseen in the local Universe, hybrid sources may be much  more prevalent in the early Universe \citep[e.g.][]{Mao10}. Consequently, understanding the formation of sources such as J1649+2635 provides us with an excellent probe for the higher-redshift Universe. 

Here we offer three possible causes for the unusual association of a spiral galaxy with large-scale double-lobed radio emission. 

\begin{itemize}
\item{Perhaps J1649+2635 is in the unique position of having undergone a planar minor merger that would not introduce angular momentum perpendicular to the plane, so as to not disrupt the spiral structure too much \citep{Naab06}. The merger would have to dump sufficient matter ($\sim6.6 - 15.5$\,M$_{\odot}$\,year$^{-1}$, assuming 0.1 efficiency) onto the central black hole so as to trigger radio lobes from the AGN (perhaps the black hole spins very rapidly similar to the \citet{Bagchi14} model), and it is perhaps this process that quenches the star-formation giving this source its red colour over a few Gyr. The presence of the large diffuse optical halo makes this process appear unlikely as the diffuse halo suggests multiple mergers have occurred, whereupon a spiral structure cannot be sustained \citep[e.g][]{Bournaud07}.}
\item{Alternatively, J1649+2635 may be a ``normal'' elliptical galaxy that has undergone some event (such as a merger) that has caused it to grow spiral arms. However, spiral arms require large reservoirs of gas to maintain. The red colour of this galaxy suggests that the gas supply has been cut-off, thus the spiral arms may decay quite quickly. }
\item{We offer one final possible explanation for this unique source. Seyfert galaxies are known to harbour AGN, and yet are predominately disk or spiral galaxies. Seyfert galaxies are also predominately found in the field. What happens if a Seyfert galaxy is in a dense environment? Could the IGM support large-scale radio lobes? }
\end{itemize}

\section{ Conclusions}
In this paper we have presented the best example to date of a grand-design spiral hosting a kpc-scale double-lobed radio source. 

In many ways, J1649+2635 is a ``normal'' red spiral galaxy. It has clearly visible spiral arms and infrared data from WISE and NUV data from GALEX both suggest that there is ongoing star-formation with SFR\,$\sim$ 0.26 -- 2.6\,M$_{\odot}$\,year$^{-1}$. J1649+2635 appears to reside in a group environment (M$_{dyn} = 7.7^{+7.9}_{-4.3} \times 10^{13}$\,M$_{\odot}$), which is consistent with red spirals. 

On the other hand, J1649+2635 hosts a $\sim 85$\,kpc double-lobed radio source. The association of large double-lobed radio sources with spiral galaxies is exceedingly rare in the local Universe. Moreover, the galaxy hosts a $\sim$96\,kpc diffuse optical halo whose integrated surface brightness is comparable to that of the spiral disk. The halo is similar to those found around the centrally dominant cD galaxies in massive clusters. This is the first instance of a grand-design spiral hosting a large diffuse optical halo. 

The presence of a large and diffuse optical halo and the overdensity of galaxies detected by SDSS implies that J1649+2635 sits in a group-sized gravitational halo. The non-Gaussian redshift histogram supports the theory that the environment is unrelaxed, hinting at a merging system or unvirialised group. The black hole mass that we estimate for this source, M$_{\rm BH} \sim$ 3--7 $\times$ 10$^8$\,M$_{\odot}$, is consistent with the galaxy having perhaps undergone a merger.

We considered three possible causes for the unusual association of a spiral galaxy with large-scale double-lobed radio emission. We suggested a planar merger may be responsible for the red colour and black-hole mass. However, if the diffuse halo is indicative of multiple mergers, this theory seems unlikely. We considered whether J1649+2635 is a ``normal'' elliptical galaxy that is undergoing a merger and serendipitously appears spiral for a brief period of time. Finally, we speculate whether J1649+2635 may be a Seyfert galaxy that is in a denser-than-normal environment. 

Further observations are required to test these predictions. For example, long-slit or integrated field spectroscopic data over the source would enable us to determine the stellar population in the spiral arms. If large velocity dispersions are observed, this may be evidence for a recent merger. Multi-object spectroscopy over the field would provide us the larger-scale clustering environment. Observations of the HI-21cm line in the system would enable us to determine the atomic gas mass and kinematics of this source and hence deduce if the HI properties are similar to other spiral galaxies. High-resolution radio imaging of the AGN will enable us to probe the parsec-scale structure at the black hole and gain insight into the AGN activity. Finally, to determine whether the large-scale diffuse radio emission from J1649+2635 is consistent with a re-activated radio source, similar to Speca and J2345+0449, observations of the diffuse radio emission are required. 

With the advent of citizen science projects such as Galaxy Zoo and Radio Galaxy Zoo, perhaps many more of these enigmatic objects may be discovered. With this in mind, we propose calling these sources Spiral DRAGNs, the DRAGNs acronym deriving from Double-lobed Radio sources Associated with Galactic Nuclei \citep{Leahy93}.

\section*{Acknowledgements}
The National Radio Astronomy Observatory is a facility of the National
Science Foundation operated under cooperative agreement by Associated
Universities, Inc. RD was a summer student at the National Radio
Astronomy Observatory. The portion of this research for which TM is
responsible was performed while he held a National Research Council
Research Associateship Award at the Naval Research Laboratory. These
results made use of the Discovery Channel Telescope at Lowell
Observatory.  Lowell is a private, non-profit institution dedicated to
astrophysical research and public appreciation of astronomy and
operates the DCT in partnership with Boston University, the University
of Maryland, the University of Toledo and Northern Arizona
University. The NSF funded the construction of the LMI under
grant AST-1005313.

We are grateful to the MNRAS editors and the Referee for their helpful
feedback, and to Jay Blanchard, Chris Carilli, Phil Hopkins, Huib
Intema, Kristina Nyland, Tom Pannuti, Anna Scaife, Lor\'{a}nt
Sjouwerman, Jamie Stevens, Gustaaf van Moorsel, Lisa Young and many
others for numerous useful and insightful discussions.

\clearpage

\label{lastpage}
\end{document}